\documentclass[12pt]{article}
\usepackage{times}
\usepackage{graphicx}
\usepackage{epsfig}
\usepackage{wrapfig}
\usepackage{pictexwd}
\usepackage{multicol}
\usepackage{color}
\definecolor{orange}{rgb}{1,.6,0}
\definecolor{purple}{rgb}{.8,0,1}

\newdimen\psize
\newdimen\tdim
\tdim=\unitlength
\def\stpltsmbl{\setplotsymbol ({\small .})}

\newbox\phru
\setbox\phru=\hbox{\beginpicture
\setcoordinatesystem units <\tdim,\tdim>
\stpltsmbl
\setquadratic
\plot
0 0
2.5 3
5 0
7.5 -3
10 0
/
\endpicture}
\def\photonru #1 #2 *#3 /{\multiput {\copy\phru}  at
#1 #2 *#3 10 0 /}
\newbox\phrd
\setbox\phrd=\hbox{\beginpicture
\setcoordinatesystem units <\tdim,\tdim>
\stpltsmbl
\setquadratic
\plot
0 0
2.5 -3
5 0
7.5 3
10 0
/
\endpicture}
\def\photonrd #1 #2 *#3 /{\multiput {\copy\phrd}  at
#1 #2 *#3 10 0 /}
\newbox\phdr
\setbox\phdr=\hbox{\beginpicture
\setcoordinatesystem units <\tdim,\tdim>
\stpltsmbl
\setquadratic
\plot
0 0
3 -2.5
0 -5
-3 -7.5
0 -10
/
\endpicture}
\def\photondr #1 #2 *#3 /{\multiput {\copy\phdr}  at
#1 #2 *#3 0 -10 /}
\newbox\phdl
\setbox\phdl=\hbox{\beginpicture
\setcoordinatesystem units <\tdim,\tdim>
\stpltsmbl
\setquadratic
\plot
0 0
-3 -2.5
0 -5
3 -7.5
0 -10
/
\endpicture}
\def\photondl #1 #2 *#3 /{\multiput {\copy\phdl}  at
#1 #2 *#3 0 -10 /}
\newcommand{\cmt}[1]{{#1}}
\newcommand{\uncmt}[1]{#1}
\newcommand{\mynewpage}{\bigskip\bigskip\par}

\raggedright
\begin{document}

\begin{centering}
~\\
\rule{0pt}{10pt}\\
{\Huge Sidney Coleman's Harvard}\\
\rule{0pt}{10pt}\\
{\Large 
A talk presented at the April 2016\\
APS meeting in Salt Lake City\\
Session: Sidney Coleman Remembered\\
 - Correspondence and Commentary\\
\rule{0pt}{10pt}\\
Howard Georgi\\
Harvard\\}
~ \\ 
~~~~~~\parbox{.9\hsize}{\raggedright\large Abstract: The speaker had the great
  good fortune to 
  take an undergraduate course in group theory from Sidney Coleman, and
  (after 
graduate school away) was hired by Coleman to a postdoctoral position and
eventually became a faculty colleague. He will 
share some still vivid memories of this remarkable character. \par}
\end{centering}

\newpage\normalsize

\uncmt{\mynewpage
$$\includegraphics[height=.6\hsize]{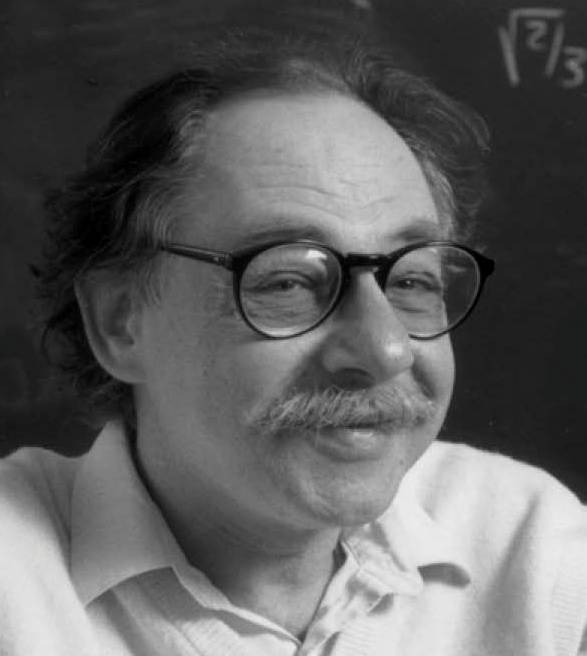}$$ }

\cmt{
If some crucial deadline kept you working to the wee hours of the morning
in your office in the Harvard Physics buildings in the late '60s or the '70s, you were
likely to see a silent, solitary figure wandering the halls, deep in thought.  This
was not the ghost of Albert Einstein (although there was an uncanny
resemblance).  It was Sidney Coleman, one of the great minds and characters
in particle physics and quantum field theory.  With his remarkable
intellect and his unique persona, Sidney put his personal stamp on
theoretical physics at Harvard for decades. 

I hope to give some impressions of what it was like to be at Harvard
  with Sidney Coleman in his prime.  I will not spend much time going over the
  wonderful physics he did, but will focus on Sidney himself. 
 I will introduce myself gradually as I go
  along, but I did want to start with a disclaimer.  
}

\subsubsection*{Disclaimer}
\uncmt{\mynewpage
$$\includegraphics[height=.6\hsize]{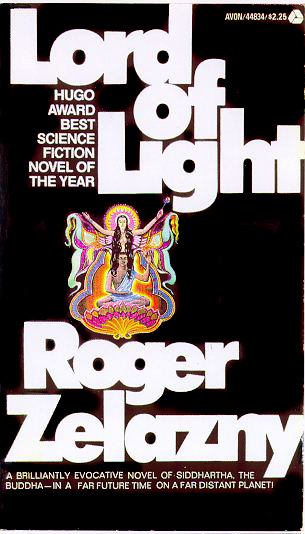}$$ }

\cmt{I don't pretend to any objectivity.  I loved Sidney.  \\
Sidney taught me my first formal course in group theory, and it was also
the first well-taught course I had in the Harvard physics department.\\
Sidney hired me as a post doc.\\
Sidney introduced me to my favorite book (Lord of Light by Roger Zelazny).\\
Sidney supported my tenure case, and that was in spite of having to give
evidence in front of the Harvard President in a zombie like state at 9am. 

~

I affirm that these
  impressions will be an accurate representation of what I remember from
  this period. But  I cannot guarantee that my memory is 100\%
  accurate, and I have not had the time nor would I have the competence to
do all the historical research necessary to get the details 
100\% right.  I am sure that some of my dates
are a litle bit off.  And I apologize to people I have left out or
misrepresented.  I hope that you will set the record straight.
}

\subsubsection*{1964}
\uncmt{\mynewpage
$$
\beginpicture
\setcoordinatesystem units <\tdim,\tdim>
\put {\includegraphics[height=.8\psize]{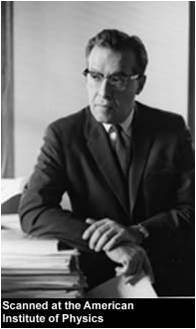}} at 0 0
\endpicture
$$ }

\cmt{I arrived at Harvard as a freshman in 1964, knowing that I wanted to
  do theoretical physics, chemistry or math.  A wonderful chemistry course
  taught by E. Bright Wilson soon convinced me that what I was really interested
  in was physics.

~

My naive freshman impression when I arrived
was that the Harvard Theory
group was led by Julian Schwinger.
This seemed reasonable since Schwinger shared the Nobel prize in 1965 while I
was an undergraduate.

~

But in fact, Julian was on his way out.  He was doing source theory 
and before long he would retire
to California.

}

\subsubsection*{The young Coleman}
\uncmt{\mynewpage
$$
\includegraphics[height=.45\psize]{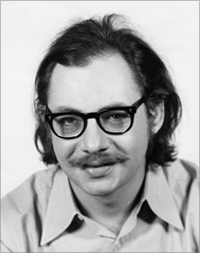}
$$ }

\cmt{The real intellectual
leader of the group when I arrived was
the young Sidney Coleman
(photo https://www.aip.org/history-programs/niels-bohr-library/oral-histories/31234)
~

Sidney arrived at Harvard in the early '60s, riding a wave of new ideas about
the application of approximate symmetry arguments to particle physics.
With his thesis advisor Murray Gell-Mann at Caltech, 
his good friend Shelly Glashow, and others, he showed the particle physics 
community how to calculate many
measurable properties of strongly interacting particles using the algebraic
techniques of group representations for continuous groups like $SU(3)$. 

~

Sidney was a wonderful teacher.  To this day, there are things I find very
confusing about group theory
that I remember thinking I understood clearly while I was under
the spell of Sidney's voice in the course I took from him over 50 years ago.
}

\subsubsection*{Shelly}
\uncmt{\mynewpage
$$
\beginpicture
\setcoordinatesystem units <\tdim,\tdim>
\put {\includegraphics[height=.45\psize]{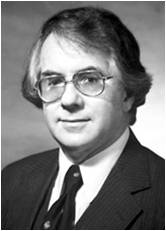}} at 0 0
\endpicture
$$ }

\cmt{Shelly Glashow, a Schwinger student who would become my close collaborator
and close friend a few years later, 
arrived at Harvard as a Professor a bit later in 1966
--- he was Sidney's mentor in many ways and he and Sidney were collaborating
on various things --- but I suspect
that even then, as in 1971, when I
returned as a postdoc, it was Sidney
who chose the menu at the Peking
on Mystic after the seminars.  I actually attempted to organize an informal
reading course with Shelly in my last year as an undergraduate.  It is
probably good that this didn't work.  I am pretty sure that I was NOT ready to
deal with Shelly back in 1967.  Fortunately, the Harvard general education
rules forced me to 
take a course in a different area, and I choose a course in abnormal
psychology instead. 

~

It was not until later that I saw Sidney and Shelly together.

}

\subsubsection*{Yale}
\uncmt{\mynewpage
$$
\beginpicture
\setcoordinatesystem units <\tdim,\tdim>
\put {\includegraphics[height=.45\psize]{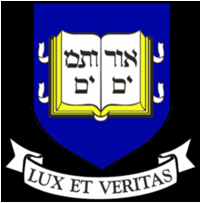}} at 0 0
\endpicture
\quad\quad
\beginpicture
\setcoordinatesystem units <\tdim,\tdim>
\put {\includegraphics[height=.45\psize]{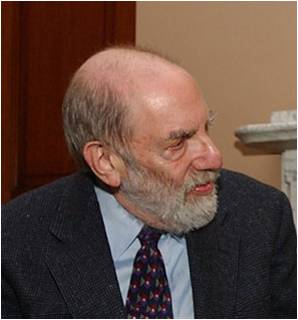}} at 0 0
\endpicture
$$ }

\cmt{In 1967, I went off to Yale grad school (because my fianc\'e was at Vassar). 
There, I worked with another Schwinger student, Charlie
Sommerfield.

~

Soon after I arrived at Yale in 1967, I saw Weinberg's ``Model of Leptons''
paper.
Like most everyone else (including, I think, Steve himself), I ignored it
because it didn't look renormalizable  to me and I didn't know what to make
of it. 
I assume that history will eventually annoint 1967 as the birth of the
standard model.  But that is not what it felt like at the time.  
The period from '67 to '71
was just confused!

 ~

In spite of the confusion,
I had great fun at Yale.  Yale is a great University where I felt at home.
At the time the particle theory  group was somewhat sleepy ---
this was
long before Tom Appelquist energized the place.  But there were lots of brilliant
and very friendly people like Charlie, Feza Gursey, and Sam
MacDowell and I learned a lot.  However,
I was very excited when in my final year at Yale, 
Sidney called me to offer me a postdoc position back at Harvard.}

\subsubsection*{Delicious confusion}
\uncmt{\mynewpage
$$\includegraphics[height=.6\hsize]{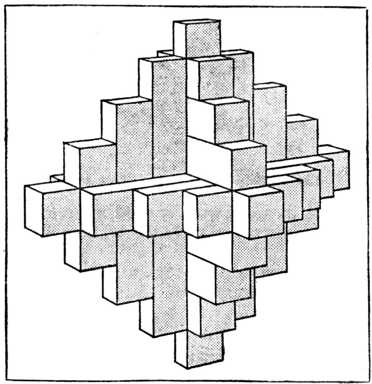}$$ }

\cmt{I am not a fan of anthropic reasoning, but I do often think that
God got up one morning 13.8 billion years ago and she thought to herself, ``all
these universes I have been creating are pretty boring.  Today I am going
to build one in which the rules are like a Chinese puzzle, a
multidimensional jig-saw puzzle in which the pieces are all only subtly
different from one another but
fit with each other in crazy, intricate ways.  I will
sprinkle it with clues but not make anything obvious.  It will be such
delicious fun!''

~

I think our world of particle physics 
is like that.  The Lie Algrebras that Sidney and I both loved 
are everywhere, but in several different
guises and for lots of different reasons. Some are
approximate and look accidental while other are dynamical and look
fundamental. Some are broken, some are not. It is crazy.  It is confusing.

}

\subsubsection*{Returning to Harvard in 1971 as the dam burst
}
\uncmt{\mynewpage
$$
\beginpicture
\setcoordinatesystem units <\tdim,\tdim>
\put {\includegraphics[height=.45\psize]{colemans.jpg}} at 0 0
\endpicture
\quad\quad
\beginpicture
\setcoordinatesystem units <\tdim,\tdim>
\put {\includegraphics[height=.45\psize]{shelly.jpg}} at 0 0
\endpicture
$$ }

\cmt{I returned to Harvard at exactly the right time in the fall of 1971. 
When Gerard 't Hooft  and Martinus Veltman  and
others finally figured this out in the early 1970s, the floodgates opened
because quantum field theorists had a huge new world of theories that they
suddenly had the tools to explore. At the same time, experimental particle
physicists were pushing their machines beyond the 1 GeV energy scale and
beginning to see evidence of new and surprising physics at (what we then
thought of as) high energy. The next few years were a remarkable confluence
of theoretical and experimental progress in particle physics.  

~

I joined Sidney and Shelly now as a colleague, and it was a revelation.
This may have been the beginning of my fascination with how the minds of
great physicists work.  Sidney said in that period that he and Shelly had
``twin minds.''   I believe that what he meant by that was that their first
instinct was to turn any problem into a symmetry argument.  That made sense
to me because my instinct is just the same.  And neither of them was a
system-builder.  Unless they were collaborating with others who worked
differently, they seldom wrote back-to-back series of papers. 
But beyond that, I think that
their minds could not have been more different.  
In an APS Oral History Interview Sidney said ``the kind of physics I like
to do is more like guerilla 
warfare than an expedition. I like to find the problem, solve it,
preferably in some snappy, elegant, striking way and then go on to the next
problem.''  But if Sidney was a guerilla, Shelly is a terrorist!  He is not
completely happy unless he is toppling the conventional wisdom. 

}

\subsubsection*{Youngsters}
\uncmt{\mynewpage
$$
\beginpicture
\setcoordinatesystem units <\tdim,\tdim>
\put {\includegraphics[height=.4\psize]{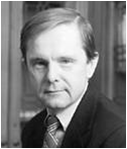}} at 0 0
\endpicture
\quad\quad
\beginpicture
\setcoordinatesystem units <\tdim,\tdim>
\put {\includegraphics[height=.4\psize]{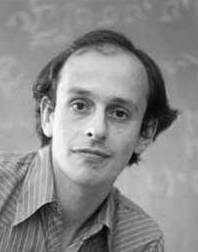}} at 0 0
\endpicture
\quad\quad
\beginpicture
\setcoordinatesystem units <\tdim,\tdim>
\put {\includegraphics[height=.4\psize]{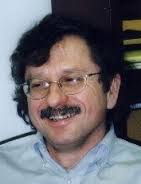}} at 0 0
\endpicture
$$ 
$$
\beginpicture
\setcoordinatesystem units <\tdim,\tdim>
\put {\includegraphics[height=.4\psize]{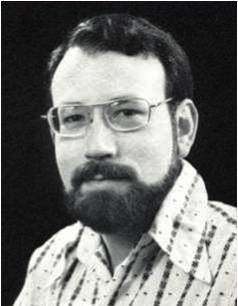}} at 0 0
\endpicture
\quad\quad
\beginpicture
\setcoordinatesystem units <\tdim,\tdim>
\put {\includegraphics[height=.4\psize]{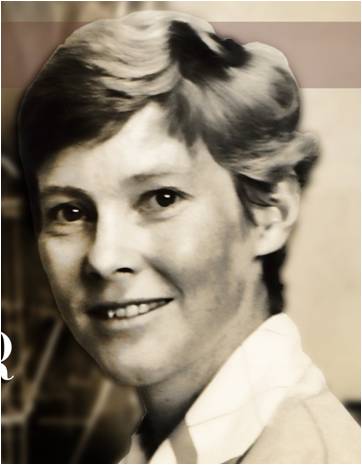}} at 0 0
\endpicture
\quad\quad
\beginpicture
\setcoordinatesystem units <\tdim,\tdim>
\put {\includegraphics[height=.4\psize]{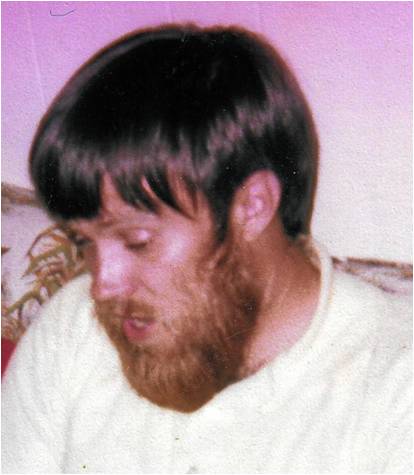}} at 0 0
\endpicture
$$ }

\cmt{There I also found a wonderful
  cast of junior characters.
Tom Appelquist was a junior faculty member.  David Politzer and Erick
  Weinberg were students of Sidney.  Joel Primack was a Junior Fellow at
  Harvard from 1970-73. A Stanford product, he was an excellent field
  theorist and a spectacular wine steward --- I can still dimly remember
the taste of the Latour '64
  he imported for the Society of Fellows. 
}

\subsubsection*{Sam Ting}
\uncmt{\mynewpage
$$\includegraphics[height=.6\hsize]{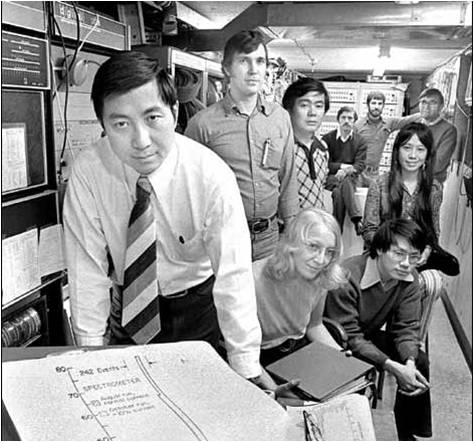}$$ }

\cmt{Initially, Helen Quinn was just hanging out.  She was there with her
  husband Dan who was working with Sam Ting. Sidney and Shelly immediately
  recognized a gift and signed her up as an honorary research fellow and
  she moved into a faculty position the following year.   

~

It was wonderful fun talking to Helen for many reasons.  She talked about
the dearth of women in physics and the different culture in which she was
raised in Australia. And 
without giving away any secrets, She gave us a bit of a sense of what
life in Sam Ting's group was like. This made the bizarre bicoastal November
revolution, when it came, a bit more believable.

}

\subsubsection*{Alvaro}
\uncmt{\mynewpage
$$\includegraphics[height=.8\psize]{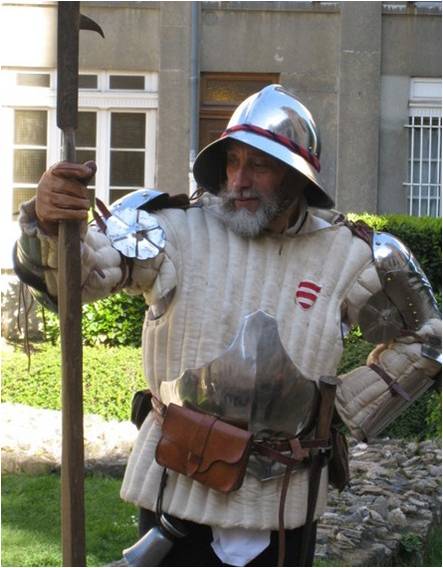}$$ }

\cmt{I later
learned that I was Harvard's 7th choice for a postdoctoral position 
--- the first 6 having gone elsewhere.

~

I think I learned this from Alvaro DeRujula, who arrived as a postdoc in
'72 the
year after I did.  He had been even further down the list than I was, and
told an amusing story about the rejection letter he received from
Sidney. He said that the letter was so polite and complimentary to him that by
the end he felt sorry for Harvard for not accepting him.

}

\subsubsection*{Ben Lee}
\uncmt{\mynewpage
$$
\beginpicture
\setcoordinatesystem units <\tdim,\tdim>
\put {\includegraphics[height=.35\psize]{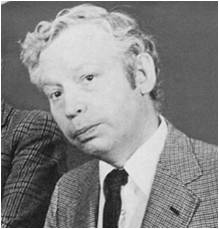}} at 0 0
\endpicture
\quad\quad
\beginpicture
\setcoordinatesystem units <\tdim,\tdim>
\put {\includegraphics[height=.35\psize]{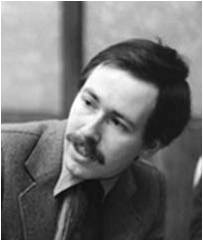}} at 0 0
\endpicture
$$
$$
\includegraphics[height=.5\psize]{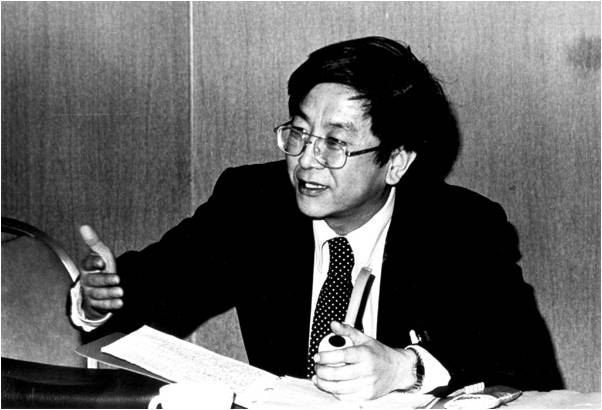}
$$ }

\cmt{The short
  version of what happened in '71 is that Gerard 't Hooft figured out how to make sense of
  spontaneously broken non-abelian gauge theories in general and Weinberg's
  model of leptons in particular. 

~

But at Harvard, we were convinced that something really important had
happened in large part by
Ben Lee, the great Korean-American particle theorist who went on to
become Director of the theory group at Fermilab before his tragic death in an
automobile accident. Ben had been trying to understand t'Hooft's
papers on the renormalizability of spontaneously broken gauge
field theories. He had done important work on the renormalization of field
theories with spontaneously broken global symmetries, and was in a good
position to make sense out of what t'Hooft had done. He reviewed t'Hooft's
arguments for us, and emphasized the connection with Steve Weinberg's '67
paper. Ben got our attention by discussing a spontaneously
broken $U(1)$ gauge theory with the gauge boson coupled to a non-conserved
current. He gave some examples of what seemed to us at the time as miraculous
cancellations required to allow renormalization. This energized
everyone in the Harvard theory group.

~

We all dropped what we were doing and started working on various aspects of
spontaneously broken gauge theories. Tom Appelquist, Helen Quinn and Joel
Primack started work on renormalization in unitary gauge, ultimately
related to the Appelquist-Carrazone theorem.  Shelly and I started building
models.  Sidney Coleman and
his students Eric Weinberg and David Politzer started the work that led to the
famous Coleman-Weinberg paper which included the idea of dimensional
transmutation and to asymptotic
freedom. 

}

\subsubsection*{Radiative Corrections as the Origin of Spontaneous Symmetry Breaking
}
\uncmt{\mynewpage
$$\includegraphics[height=.5\hsize]{erick-w2.jpg}$$ }

\cmt{While Coleman's contributions to the enormous progress made in particle
theory in the '70s were huge, he was usually not directly involved in
interpreting the exciting experimental results. But he was always among the
first to understand new theoretical ideas, often much more clearly than the
inventors themselves. He was often the first to put new theoretical ideas
on a firm footing and to understand their connection with deep issues in
the foundations of physics. And he frequently took the lead in explaining
them clearly to the community.  

~

It was characteristic of Coleman that many of his deepest and most
important contributions are hidden in long papers that might seem to the
casual observer to be purely technical, working out of some minor
mathematical detail. Two wonderful examples of this from the 1970s are the
papers "Radiative Corrections as the Origin of Spontaneous Symmetry
Breaking" (Coleman and Weinberg, 1973) and "Quantum sine-Gordon Equation as
the Massive Thirring Model" (Coleman, 1975). In the first of these Coleman
and his student Erick Weinberg solve a puzzle. They begin like this,  

~

{\it Massless scalar electrodynamics, the theory of the electromagnetic
interactions of a mass-zero charged scalar field, has had a bad name for a
long time now; the attempt to interpret this theory consistently has led to
endless paradoxes. In this paper we describe how nature avoids these
paradoxes: Massless scalar electrodynamics does not remain massless, nor
does it remain electrodynamics. }

~

It might sound from this introduction that they are simply giving 
a consistent account of a
pathological theory, but the paper was MUCH more than that. 
It was enormously influential as a handbook for
dealing with scale violation in quantum field theory. Coleman had been
thinking hard about scale invariance since the late 1960s. In this paper,
written soon after the revolution of spontaneously broken non-Abelian gauge
theories, Coleman and Erick Weinberg pulled together all the most useful
techniques and described them with Sidney's characteristic clarity. In the
process they discovered an important and very general phenomenon. They say, 

~

{\it The surprising thing is that we have traded a dimensionless parameter,
$\alpha$, on which physical quantities can depend in a complicated way, for
a dimensional one --- on which physical quantities must depend in a trivial
way, governed by dimensional analysis. We call this phenomenon dimensional
transmutation.}

~

We now know that dimensional transmutation is responsible for many of the
surprising features of the strong interactions at high energies that were
appearing in experiments when their paper was written.  
}

\subsubsection*{The method of the virtual guru
}
\uncmt{\mynewpage
$$\includegraphics[height=.6\hsize]{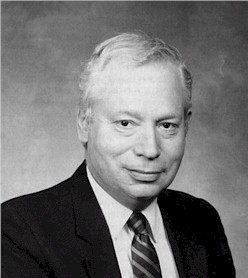}$$ }

\cmt{Steve Weinberg himself made the switch from MIT to Harvard in 1973 .  Over
  the next few years I had the fun of getting to know him and collaborating
  with another of the real greats in the field. As I had been with Sidney and Shelly, I
  was astonished at how differently his mind worked.  Steve seemed to work
  backward from the general case to specific examples.  And unlike Sidney, he
  was very much into elaborate ``expeditions'', and if Shelly is a
  terrorist, Steve is more of a Royalist. 

I think it is one of the great ironies that Steve got the Nobel Prize for a
specific model, because that is really not his style.
 I found this really
  bizarre, but very useful.  His thinking was so
foreign to me that we often approached the same problems from opposite
ends, and that was sometimes very useful because we got to a deeper
understanding.  

~

Sidney's student David Politzer used to talk about ``the method of the virtual guru'' and
this applied very well to Steve.  The idea was that if you studied how
great physicists think, and you came upon a problem that you thought they could do
better than you, you could adopt their methods and make progress.  I am not
sure that I could use Sidney or Shelly as examples this way, because our
approaches tended to be a bit similar anyway.  But Steve was a terrific
virtual guru, because his very systematic and general approach to problems
was something I could try to apply, even though I wasn't anywhere near as
good at it as Steve was.

~

But also, along with Sidney and Shelly, Steve was the third out of the 
three physicists whom I
had gotten to know really well who were universally regarded as great. And
I was convinced that each of them occupied a completely different
region in the space of intelligence.  This has come to be very important to
me in my thinking about diversity.  
I have come to believe that the number of ways of being a great scientist
is at least as large as the 
number of great scientists. Because of these observations, I strongly
suspect that there are many other 
ways of being a great scientist that we have not yet seen and that we may
never see unless we open up the scientific enterprise to people with very
different backgrounds. 
This is one reason that I believe diversity in science is so important.
It is also a reminder of how
silly it is to rank scientists (or students for that matter) by any small
number of measures.  
}

\subsubsection*{SLAC}
\uncmt{\mynewpage
$$\includegraphics[height=.8\psize]{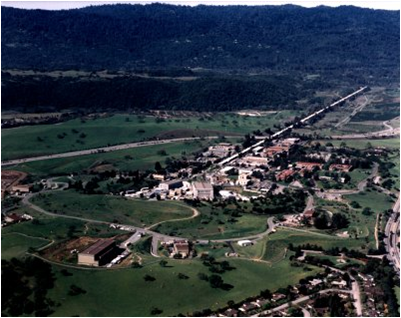}$$ }

\cmt{On the experimental front, after years without easily interpretable
  dynamical information about the strong interactions, deep inelastic
  scattering  at SLAC was
  beginning to make sense and to look really interesting (though still quite confusing).

~

It is really amazing to think back on those days and remember how different our
day-to-day activities were from what we do today.  We were programing with
fortran and punch cards.  I remember clearly when we got our first
electronic calculators - first Wangs -  and thereafter Hewlett-Packards.
We wrote our papers long-hand on yellow sheets of paper
and handed them to poor martyred secretaries
who typed them on IBM selectric typewriters.  You felt terrible about
making even a tiny change after the initial typing job.  Donald Knuth
should probably get the Nobel Peace Prize.

~

We sent out preprints and had stacks of preprinted  post cards to send to
other physicists asking for their papers.  These had boxes that you just
filled in --- ``I would appreciate a copy of your paper BLANK - which
appeared in BLANK -  After I had written my first paper as a graduate
student in 1970 with Postdoc John Rawls, I received one of these from
Lowell Brown - which I have reproduced here: 
}

\subsubsection*{Preprint request}
\uncmt{\mynewpage
$$\beginpicture
\setcoordinatesystem units <.85\hsize,.55\hsize>
\linethickness=4pt
\put {\large Dear Dr. $\underline{~~~{Georgi}~~~~~}$} [l]  at .1 .85
\put {\large I would be very grateful if you could send me a } [l] at .15 .75
\put {\large copy of your paper:} [l] at .1 .65
\put {\large
$\underline{{Anomalies\;of\;the\;Axial-Vector\;Current\;\cdots}}$}
[l]  at .1 .55  
\put {\large which appeared in} [l] at .1 .45
\put {\large
$\underline{{my\;notes\;in\;1968\;I\;suspect\rule{19ex}{0pt}}}$}
[l]  at .1 .35
\put {\large Professor Lowell Brown, Physics Dept} [l] at .1 .25
\put {\large University of Washington, Seattle, WA 98195} [l] at .1 .15
\putrule from 0 0 to 0 1
\putrule from 1 0 to 1 1
\putrule from 0 0 to 1 0
\putrule from 0 1 to 1 1
\endpicture$$ }%

\cmt{By the way, I knew Lowell from his time at Yale and I think the he knew that I
  would amused rather than annoyed.
 }

\subsubsection*{The graduate student revolt}
\uncmt{\mynewpage
$$\includegraphics[height=.55\hsize]{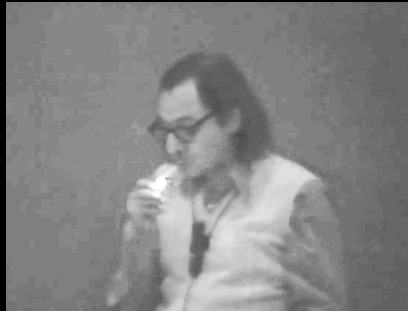}$$ }

\cmt{Sidney professed not to love teaching, even the teaching of graduate
  students, except for extraordinary students like Erick and David Poltizer
  who could be treated as colleagues almost from the beginning.
He claimed that is was just a job and that he did it well just for the
satisfaction of a job well-done.  But he certainly did it well.  His
Quantum Field Theory lectures from the '75-'76 academic year
 were recorded and are available on
the Harvard Physics Department Video Archives.  The photo shows another big
change from those days --- Sidney was lighting up a cigarette in class.

~

But while Sidney professed no love of teaching,  he tried to be fair.
In the late '70s, 
when a group of graduate students complained that they were not getting
enough contact with the faculty beyond their advisors.  Sidney listened and
got the point.
In fact, the Harvard system was a bit of a hold-over from the the old days
of Julian Schwinger, who had many students to whom he gave very interesting
problems, but who had very little organized contact with the department
except through courses.  To deal with this problem, we organized a series
of informal seminars given by the students and postdocs with the faculty in
attendance.  This was a great system (so obvious that I suspect that
many departments were already doing it), and it has survived  almost
continuously with just a few
alterations to this day.

}

\subsubsection*{The junior-faculty revolt}
\uncmt{\mynewpage
$$\includegraphics[height=.55\hsize]{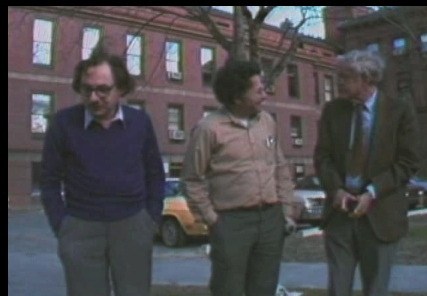}$$ }

\cmt{The junior-faculty revolt was similar.  In those day, Harvard did not
  treat its Junior faculty very well.  In some groups, they were glorified
  post-docs, with a higher salary but significant teaching
  responsibilities.  Very few got tenure.  We got away with this for a long
  while only because Harvard was a place where they could do great research, and
end up in a great position for their next job.  We do much better now, and
actually have a tenure track system that means something.

~

Anyway, in the early '80s, the junior faculty noticed that not only were
they on a very precarious track to tenure,  but they 
always seemed to end up teaching difficult undergraduate courses, rather
than graduate courses in their specialties.  

Again, Sidney stepped up.  He volunteered to take his turns teaching the
beginning classes.  This was a huge commitment for him, because the
beginning classes were very early in the morning - 11am or sometimes even
9:30am.  But Sidney made the supreme sacrifice with the predictable result
that he gave great courses, and left a legacy of interesting pedagogical tricks and
demonstrations some of which are still used. The picture is another one
from the department video archives, on ``The Music of the Protons'' from
1983 - ``listening'' to the precession of proton spins in the earth's
magnetic field.   This
had to be done outside away from stray magnetic fields.  With Sidney are
Ike Silvera and Ed Purcell.
}

\subsubsection*{Quantum
  sine-Gordon equation as the massive thirring model}
\uncmt{\mynewpage
$$\includegraphics[height=.6\hsize]{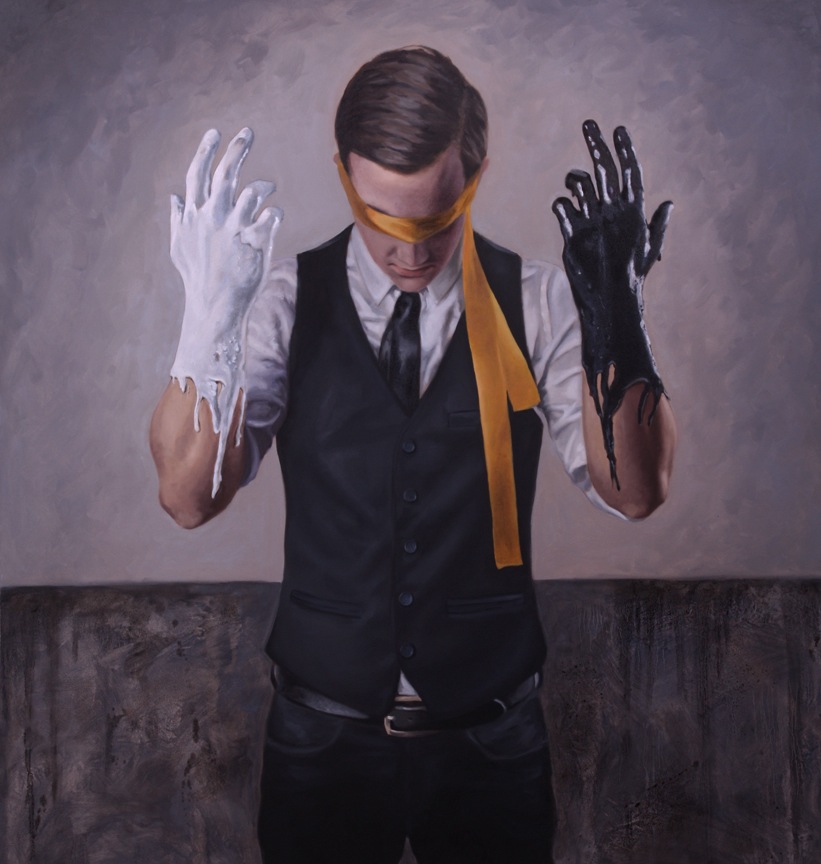}$$
}

\cmt{I wish I had a picture to show you of Sidney's nocturnal wandering.
  It wasn't just at night.  When he was really chewing on an interesting
  problem,  he often wandered around at all hours.   I remember the day (I
  think it was during the day) that he wandered in and annouced ``I am
  possessed by the spirit of Freeman Dyson!''  This was when the ideas were
  coming together for his famous paper ``Quantum
  sine-Gordon equation as the massive thirring model''.  I think that this
  was one of his very favorites because it was just such beautiful Quantum
  Field Theory.

~

Again, it is worth quoting.

{\it Thus, I am led to conjecture a form of duality for this two-dimensional theory.
A single theory has two equally valid descriptions in terms of Lagrangian field
theory: the massive Thirring model and the quantum sine-Gordon equation.
The particles which are fundamental in one description are composite in
the other. Speculation on extending these ideas to four dimensions is left
as an exercise for the reader. }
  }

\subsubsection*{Polymath}
\uncmt{\mynewpage
$$\includegraphics[width=.4\hsize]{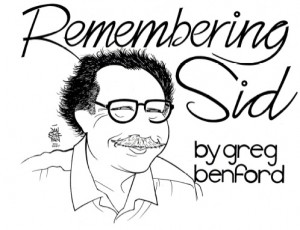}$$
}

\cmt{Having Sidney as a friend was a constant drain on the ego.   He was an
  expert on many things, not just physics.
The way Shelly explained this to me long ago was by describing going to a movie
with Sidney.  You go to the movie and are entertained and have a good time and then
afterwards, Sidney explains to you what the movie was really about, and you
feel stupid.

~

He was
particularly famous in the science fiction community -- almost as much as
in particle physics. 
He had a deep knowledge of science fiction.  For much of his
life,
he was part owner
of Advent Publising 

~

He said this about his experience as a sci-fi publisher: {\it
I am not in science fiction for money; I am in it for joy. Formally, I am a
publisher (actually, 14\% of a publisher). This is useful: it gets me on the
 mailing list of The Proceedings of the Institute for Twenty-First Century
 Studies; it is a handy topic of conversation at parties; it is a means
 whereby I meet some interesting people; it is a better hobby than
 stamp-collecting any day. From an economic standpoint, it plays a lesser
 role in my life than returning Coke bottles for refunds.}

http://ieet.org/index.php/IEET/more/benford20150226

}

\subsubsection*{Jokes}
\uncmt{\mynewpage
$$\includegraphics[height=.55\hsize]{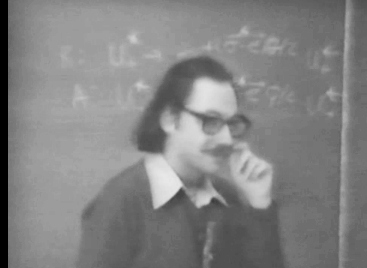}$$ }

\cmt{This last leads naturally into my last topic.
No discussion of Sidney Coleman would be complete without a discussion
of his jokes.  In his classes, Sidney was the master of the carefully
rehearsed spontaneous joke.

~

His jokes inspire a certain amount of awe, at least in me, because many of
them are so intricate that I know
I would never be able to get through them. Sidney delivered them
beautifully of course. 
But one of the things that made them funny was that Sidney thought
they were funny, and would chuckle disarmingly and twirl his moustache to
highlight the punch line.

~

I am told that there is a group working on turning his old lectures into a
book, and I hope that many of the jokes survive.

~

Here are a few from the beginning of the course.

~

The video of his quantum field theory course in 1975-76 begins with his
comment about the video equipment: {\it The apparatus you see around you is part
of a CIA survielance project.  I fall within their domain because I read
JETP Letters} [This was during the cold war and JETP is the Russian
Jounal of Experimental and Theoretical Physics]. 

~

{\it Bjorken and Drell is a very good book ... by an objective test - it is
  the
book most frequently stolen from the Physics Research Library }

~

My favorite is about notation:

~
{\it We now go to the dullest part of the lecture in which I set up my
notation. ... It will be both the dullest and the most obscure since I will
go through these things very fast because I presume that 90\% of you have
seen 90\% of what I am going to say.  Thus you will be bored 90\% of the
time and the other 10\% of the time you will be baffled, because I am
going so fast. But since it is a different 90\% and a different 10\% for
each member of the audience, there is no other way to organize it. }

\subsubsection*{Al}
Along with his carefully rehearsed jokes, Sidney also cultivated his
Einstein look. 

~

I can't resist closing by showing this wonderful caricature of Sidney produced
  by one of his sci-fi friends.

}

\uncmt{\mynewpage
$$\includegraphics[height=.65\hsize]{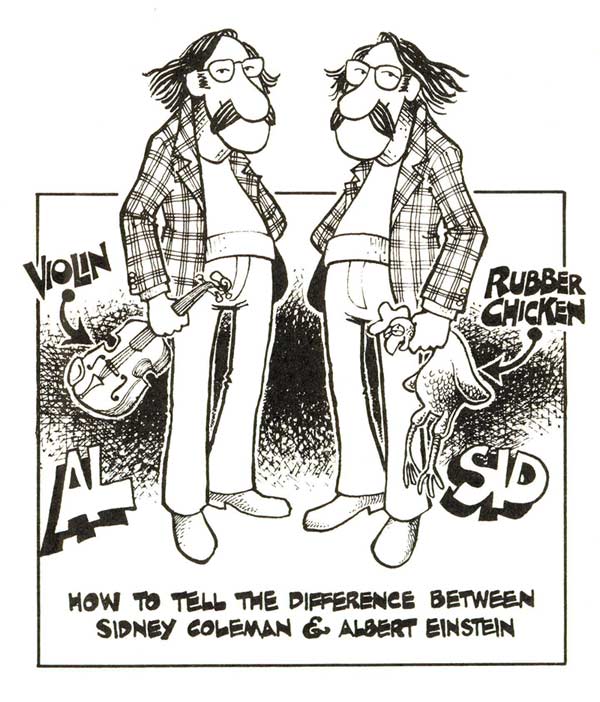}$$%
{\small Cartoon by Grant Canfield in http://efanzines.com/EK/eI36/}}

\section*{Acknowledgements}

I am very grateful to Alan Chodos for inviting me to the session.  HG is
supported in part by the National 
Science Foundation under grant PHY-1418114.

\end{document}